\documentclass[runningheads]{llncs}

\usepackage[utf8]{inputenc}

\usepackage{amsmath}
\usepackage{amssymb}
\usepackage{amsfonts}
\usepackage{color}
\usepackage{cleveref}
\usepackage[normalem]{ulem}

\usepackage[boxed]{algorithm2e}
\LinesNumbered

\usepackage{macros}
\usepackage[T1]{fontenc}
\usepackage{graphicx}

\begin{document}
\title{$\mtom$: A general method to perform various data analysis tasks from a differentially private sketch}

\titlerunning{$\mtom$}

\author{Florimond Houssiau*\inst{1} \and
Vincent Schellekens*\inst{2} \and
Antoine Chatalic\inst{3} \and
Shreyas Kumar Annamraju\inst{4} \and
Yves-Alexandre de Montjoye\inst{4}}

\authorrunning{Houssiau et al.}

\institute{The Alan Turing Institute \and UCLouvain \and MaLGa - DIBRIS, Università di Genova, Genoa, Italy \and Imperial College London\\
*These authors contributed equally.}

\maketitle

\begin{abstract}
Differential privacy is the standard privacy definition for performing analyses over sensitive data.
Yet, its privacy budget bounds the number of tasks an analyst can perform with reasonable accuracy, which makes it challenging to deploy in practice.
This can be alleviated by private sketching, where the dataset is compressed into a single noisy sketch vector which can be shared with the analysts and used to perform arbitrarily many analyses.
However, the algorithms to perform specific tasks from sketches must be developed on a case-by-case basis, which is a major impediment to their use.
In this paper, we introduce the generic \emph{moment-to-moment} ($\mtom$) method to perform a wide range of data exploration tasks from a single private sketch.
Among other things, this method can be used to estimate empirical moments of attributes, the covariance matrix, counting queries (including histograms), and regression models.
Our method treats the sketching mechanism as a black-box operation, and can thus be applied to a wide variety of sketches from the literature, widening their ranges of applications without further engineering or privacy loss, and removing some of the technical barriers to the wider adoption of sketches for data exploration under differential privacy. 
We validate our method with data exploration tasks on artificial and real-world data, and show that it can be used to reliably estimate statistics and train classification models from private sketches.

\end{abstract}

\section{Introduction}

The amount and level of detail of data collected has increased exponentially over the last two decades. Behavioral data has evolved from hand-collected medical records to GPS traces automatically recorded with a temporal resolution on the scale of seconds.
While this increased availability and precision of data has resulted in tremendous advances in research, they raise serious privacy concerns.
Modern datasets often contain highly detailed summaries of our lives, and are notoriously hard to anonymize. Individuals have indeed been shown to be easily re-identifiable in large-scale behavioral datasets, such as mobile phone metadata~\cite{de2013unique}, credit card data~\cite{de2015unique} and web browsing data~\cite{barbaro2006}.

Differential privacy (DP)~\cite{dwork2008differential}  was introduced by Dwork et al. as a property of algorithms that protect the privacy of users in a dataset.
It requires for a randomized algorithm's outputs to be distributed approximately identically whether any one individual is in the dataset or not.
The discrepancy between distributions is controlled by a parameter $\varepsilon$ known as the privacy budget.
DP is considered by many to as the gold standard definition for privacy loss in aggregated data releases.
DP mechanisms have been deployed by institutions with access to large datasets, such as Google to measure changes in mobility patterns caused by confinement measures~\cite{aktay2020google}, LinkedIn to answer analytics queries~\cite{kenthapadi2018pripearl}, and the US Census for the 2020 Census~\cite{abowd2018us}.

Most applications of DP remain limited to specialized tasks on large datasets.
Indeed, each differentially private access to a dataset consumes some privacy budget $\varepsilon$, and the total acceptable budget is fixed by the data owner for the dataset.
Once this budget has been used entirely, the dataset must be discarded.
As such, the number of accurate statistical tasks an analyst can run on a dataset is capped.
This strongly limits the utility of differential privacy in practice.
In particular, data exploration with DP is particularly challenging: it requires analysts to establish which analyses they want to perform on the dataset, and how to divide the budget between them, before accessing the data.

An increasingly popular solution to this issue is to first compute a differentially private summary of the data, called a private \emph{sketch}, which is then shared with analysts.
Once computed, the private sketch can be used as much as desired to solve new learning tasks without accessing the data anymore or using additional privacy budget. This follows from the post-processing property of DP.
Sketches have long been used as a technique to compress large-scale datasets to reduce the computational load of algorithms.
In this work, the sketch of a dataset is defined as the empirical average of some \textit{feature map} function $\featuremap$ over all records in a dataset $\dataset$: $\sketch{\dataset} = \frac{1}{|\dataset|}\sum_{x_i \in \dataset}\Phi(x_i)$.
The choice of feature map controls the specificity of the information contained in the sketch. For example, researchers have proposed sketches based on Random Fourier Features (RFF)~\cite{schellekens2019differentially} and locality-sensitive hashing~\cite{coleman2020sub} that approximate kernel density estimates of the empirical distribution.
For some specific sketches and tasks, algorithms with strong theoretical guarantees of accuracy have been developed~\cite{gribonval2017compressive}.

However, performing arbitrary data analysis tasks from sketches is difficult, as extracting the desired information from a highly compressed representation of the data is challenging. 
Each specific task and feature map $\featuremap$ would require a dedicated algorithm designed by experts.
For instance, RFF sketches have in practice only been used for a few tasks, such as Gaussian mixture modeling (GMM)~\cite{keriven2017b} or k-means~\cite{keriven2017a}.
Developing compressive methods for other data exploration tasks remains an open problem.
This is the main obstacle to using sketches for general data analysis.

\begin{figure*}
    \centering
    \includegraphics[width=0.8\linewidth, trim=0 2cm 0 2cm]{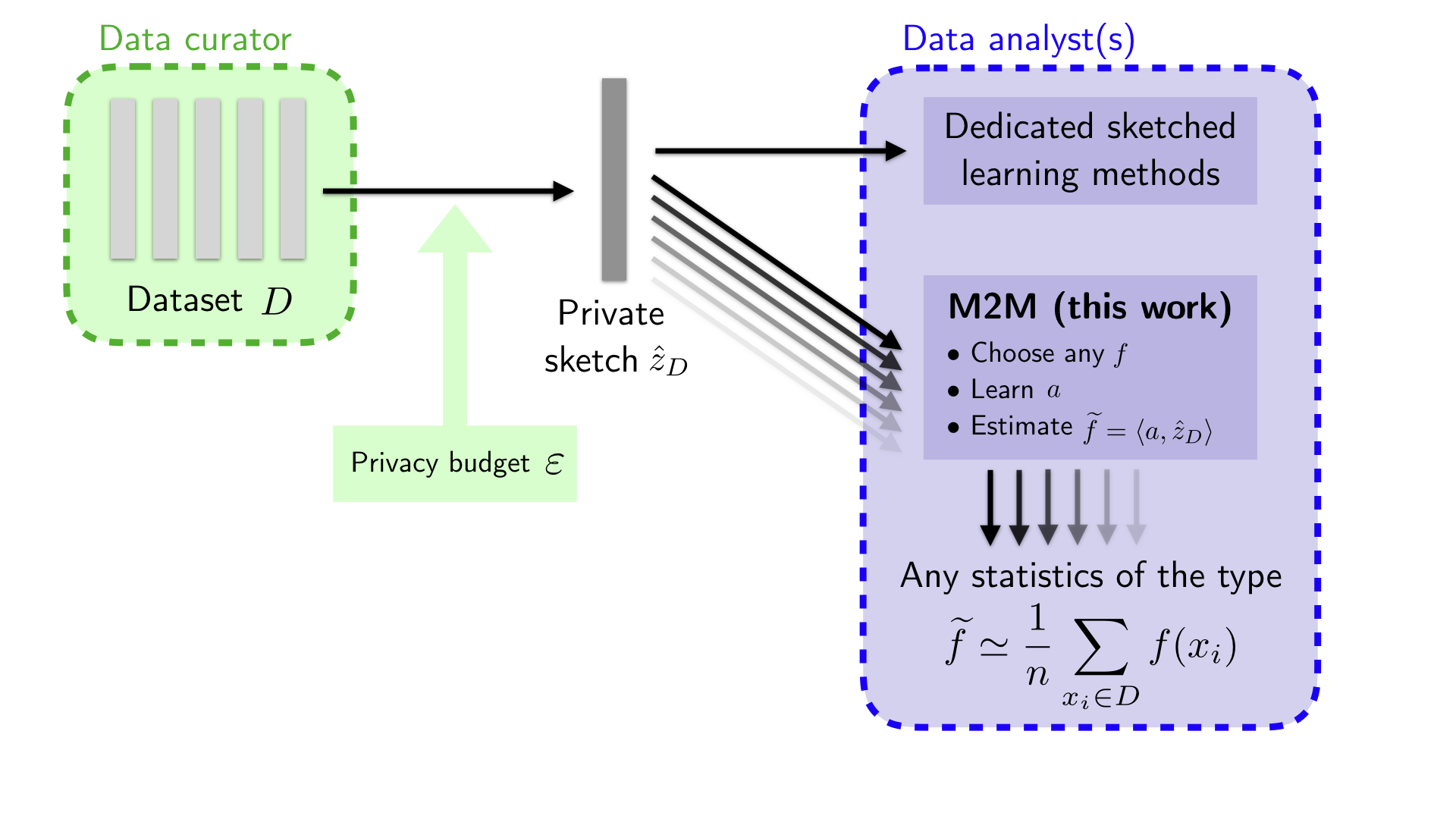}
    \caption{\label{fig:intro}\textbf{Considered setup}. The data curator releases "once and for all" a private sketch with privacy budget $\varepsilon$.
    The analyst then chooses a function $f$ and uses our $\mtom$ method to learn a vector $a \in \mathbb{R}^m$ such that $\widetilde{f} = \left<a, \privatesketch{\dataset}\right>$ approximates the empirical average value of $f$ over the dataset, $\overline{f} = \frac{1}{n} \sum_i f(x_i)$.
    This procedure can be repeated any number of times (for various choices of $f$) without additional privacy budget.}
\end{figure*}

In this paper, we introduce a \textbf{heuristic to learn from dataset sketches} as shown in \Cref{fig:intro}, which we call the moment-to-moment ($\mtom$) method. $\mtom$ allows to approximate empirical averages of functions $f$ from the sketch, $\frac{1}{|\dataset|}\sum_{x_i \in \dataset} f(x_i)$, and can in principle be applied to any feature map $\Phi$.
This method is inspired by approximation techniques for kernel methods using random features~\cite{rahimi2008random,liu2020random}. We \textbf{empirically validate our method with artificial and real-world data}, and show that a variety of tasks (moment estimation, counting queries, covariance estimation, logistic regression) can be learned from sketches with comparable performances to alternatives (synthetic data).

\section{Background}
\label{s:background}

\subsection{Sketches}
\label{section:sketched_learning}

Sketches are compressed representations of data collections, which can be used to perform some operations efficiently but approximately~\cite{cormode2012synopses,blum2020foundations}.
Sketches usually rely on randomness to achieve a compact representation size. This comes at the price of a probabilistic approximation error.
This general principle finds applications in a broad set of contexts, from data streams\cite{flajolet1985probabilistic,misra1982finding} to randomized linear algebra~\cite{drineas2006fast}.
Here, we focus on sketches that compress the dataset $\dataset = (x_1, \dots, x_n) $ to a single sketch vector $\sketch{\dataset} \in \mathbb{R}^m$ by computing the \textit{average of a nonlinear feature map} $\featuremap$, applied to each record $x_i$. 
\begin{definition}
\label{def:sketch-of-a-dataset}
Given a feature map $\featuremap: \mathbb{R}^d \rightarrow \mathbb{R}^m$, the \emph{sketch} of a dataset $\dataset = (x_1, \dots, x_n) \in \datasetset$ is
\begin{equation}
\label{eq:sketch_def}
\textstyle
    \sketch{\dataset} 
	\triangleq \frac{\Sigma_{\featuremap}(D)}{|D|} = \frac{1}{n}\sum_{i=1}^n \featuremap(x_i) \in \bb R^m,
\end{equation}
with $n=|D|$ the dataset size and $\Sigma_{\featuremap}(D)=\sum_{i=1}^n \featuremap(x_i)$ the sum of features.
\end{definition}

The representation $(\Sigma_{\featuremap}(D), |D|)$, where the sum-of-features and dataset size are distinctly encoded, is often used in practice to make it possible to further combine sketches of different datasets into a single one~\cite{cormode2012synopses}.

Typically, sketches are constructed such that scalar products approximate a specific similarity score (called \textit{kernel} $\kappa : \bb R^d \times \bb R^d \rightarrow \bb R_+$), $\langle \featuremap(x), \featuremap(x') \rangle \simeq \kappa(x,x')~\forall x, x'$~\cite{parzen1962estimation}.
This means that they can be used for kernel density estimation (KDE) , i.e. building an approximation of the data distribution $p_X$ by a density $\hat{p}(x) \triangleq \frac{1}{n}\sum_{i=1}^n \kappa(x,x_i) \approx \left<\phi(x), \sketch{\dataset}\right>$.

The feature map $\featuremap$ should be designed such that the sketch $\sketch{\dataset} $ captures enough information to solve a target learning task (i.e. the sketched KDE density $\widetilde{p}$ accurately represents the true data distribution $p_X$) while compressing the data as much as possible (i.e. the sketch size $m$ should be small).
We here review several important feature map choices.

\subsubsection{Histograms}
Histograms and contingency tables have been extensively studied in the DP literature~\cite{dwork2008differential}.
Both can be seen as a illustrative examples of sketches (in the sense of~\Cref{def:sketch-of-a-dataset}), where the feature map is
\[
\featuremap^{\HIST}(x) \triangleq \left(
I\{x \in \mathcal{P}_i\}
\right)_{i=1,\dots,m}
\in \{0,1\}^m,
\]
where $\left(\mathcal{P}_i\right)_{i=1}^m$ is a list of subsets of the data domain $\mathbb{R}^d$, and $I\{A\}$ is the indicator function which returns $1$ (resp. $0$) whenever $A$ is true (resp. false).
For 1-D histograms with $n_{bins}$ bins for example (what we call the HIST sketch), these sets are the one-dimensional bins along each component.
For this sketch, $m = d \cdot n_{bins}$.

\subsubsection{RFF Sketches}
Random Fourier Features (RFF) aim to approximate shift-invariant kernels $\kappa(x,x') = K(x-x')$.
They were initially introduced to accelerate kernel methods in machine learning~\cite{rahimi2008random}.

\begin{definition}[Random Fourier Features] 
\label{def:random-fourier-features}
Given $m' = \frac{m}{2}$ ``frequency vectors'' $\Omega = [\omega_1, \dots, \omega_{m'}] \in \mathbb{R}^{d \times m'}$ drawn $\omega_j \distiid \Lambda$, the RFF map is defined as:
$$\featuremap^{\RFF}(x) \triangleq \textstyle \left[\cos(x^T \Omega), \, \sin(x^T\Omega)\right]^T \in \mathbb{R}^m.$$
\end{definition}

The idea is that shift-invariant kernels can be decomposed as $K(x-x') = \bb E_{\omega \sim \Lambda} e^{i \omega^T (x-x')}$ where the probability distribution $\Lambda$ is the kernel Fourier transform $\Lambda(\omega) = \int K(u) e^{-i \omega^T u} \mathrm{d}u$ (owing to Bochner's theorem~\cite{Rudin1962bochnerBook}). For example, the Gaussian kernel $\kappa(x,x') = \exp\left(-\|x-x'\|^2_2/2 \sigma^2\right)$ admits a Gaussian distribution as Fourier transform, $\Lambda = \mathcal N(0, \sigma^{-2} I_d)$.
One can then show~\cite{rahimi2008random} that up to a constant scaling, $\featuremap^{\RFF}$ satisfies the kernel equation for this kernel.

RFF sketches have been successfully used for parametric density estimation tasks, such as $k-$means~\cite{keriven2017a} and Gaussian Mixture Modeling~\cite{keriven2017b}, reducing the computational resources required by orders of magnitude on large-scale datasets.

\subsubsection{RACE Sketches}
The Repeated Array-of Counts Estimator (RACE) sketch was proposed~\cite{coleman2020sub} as an alternative way to approximate KDE for so-called LSH kernels.
In RACE, the feature map $\featuremap$ takes binary values, and is constructed by concatenating $R$ independent hashing functions that each map to $W$ distinct buckets. The size of the sketch is thus $m = R\cdot W$.
RACE sketches use \textit{locally-sensitive} hash (LSH) functions: let $W \in \mathbb{N}_0$, a family $\cl H$ of hash functions $h : \bb R^d \rightarrow \{1,...,W\}$ is locally-sensitive with collision probability $\kappa$ if $\bb P_{\cl H} \left[ h(x) = h(x') \right] = \kappa(x,x')$ for all $x,x' \in \bb R^d$.

\begin{definition}[Repeated Array-of Counts Estimator] 
\label{def:race}
Given $W\in\mathbb{N}_0$, $h_j, j=1,...,R$, a set of $R = \frac{m}{W}$ hash functions drawn independently from $\cl H$, the associated RACE map is defined as:
$$\featuremap^{\RACE}(x) \triangleq \left[\iota(h_1(x))^T, ..., \iota(h_R(x))^T \right]^T \in \{0,1\}^m,$$
where $\iota : \{1,...,W\} \rightarrow \{0,1\}^W$ denotes the one-hot encoding operation.
\end{definition}

Similarly to RFF, one can show~\cite{coleman2020sub} that for all choices of LSH, there exists a kernel $\kappa$ such that the kernel equation is satisfied.

\subsection{Differential Privacy}

Differential privacy (DP)~\cite{dwork2008differential} is seen as the standard definition of privacy for aggregate data releases. It states that the distribution of a differentially private algorithm's output is similar for any two \emph{neighboring} datasets. Different relations can be considered, but in general (and for the rest of this manuscript), we consider that two datasets are neighbors if they differ by the addition or removal of any one record; this is known as “unbounded” DP\footnote{We consider only unbounded DP for conciseness, yet the private sketches from \Cref{s:dp_sketching} can be extended in a straightforward manner to the bounded DP setting. In this case no noise needs to be added to the denominator in \eqref{eq:private-sketch}.}. The guarantees of DP are characterized by a privacy ``budget'' $\varepsilon >0$ which bounds the information disclosure from the dataset.
Denote by $\datasetset$ the set of all datasets, equipped with a neighboring relation $\sim$. In this work, we consider datasets as collections of $d$-dimensional real-valued vectors $x_i \in \mathbb{R}^d$.

\begin{definition}[Differential Privacy]
\label{def:differential-privacy}
A randomized mechanism $\mechanism : \datasetset \rightarrow \mathbb{R}^m $ is $\varepsilon$-differentially private iff $\forall \dataset \sim \dataset' \in \datasetset$, $\forall S \subset \mathbb{R}^m$:
$$\proba{\mechanism(\dataset)\in S} \leq e^\varepsilon\,\proba{\mechanism(\dataset')\in S}.$$
\end{definition}

Differential privacy has several desirable properties. First, \emph{composition} guarantees that accessing the same dataset with $N$ different mechanisms respectively using budgets $\varepsilon_1, \dots, \varepsilon_N$ uses a total budget of at most $\varepsilon_{total} = \sum_{i=1}^N \varepsilon_i$.
Second, \emph{post-processing} ensures that once some quantities have been computed by a differentially private algorithm, no further operation on these quantities can weaken the privacy guarantees. 
The latter is particularly important for sketches, as it implies that all analyses ran on a $\varepsilon-$DP sketch are $\varepsilon-$differentially private.

A common method to compute a function $f$ over a dataset with $\varepsilon$-DP is the Laplace mechanism~\cite{dwork2006calibrating}. For a target function $f:\datasetset\rightarrow\mathbb{R}^m$, this mechanism adds centered Laplace noise with scale proportional to the \emph{sensitivity} of $f$.
\begin{definition}[Laplace Mechanism]
\label{def:laplace-mechanism}
The Laplace mechanism to estimate privately a function $f:\datasetset\rightarrow\mathbb{R}^m$ is defined as
$\lapmechanism_f(D) = f(D) + \xi$, where $\xi_j \sim \cl L(\beta), j = 1,...,m$ is centered Laplace noise with scale parameter $\beta = \frac{\Delta_1(f)}{\varepsilon}$. The sensitivity $\Delta_1(f)$ is defined as $\Delta_1(f) \triangleq \textstyle \sup_{D \sim D'} \|f(D) - f(D')\|_1$.
\end{definition}

\subsection{Differentially private sketching}
\label{s:dp_sketching}

We new consider privatized versions of the sketches in the form~\eqref{eq:sketch_def}.
As the considered feature maps are bounded, their sensitivities are also easily bounded; thus, the Laplace mechanism can be used to produce private versions of these sketches.
Following~\cite{chatalic2021CompressiveLearningPrivacy} we compute a sketch of the form
\begin{equation}
\textstyle
    \label{eq:private-sketch}
    \privatesketch{D} \triangleq \frac{\Sigma_{\featuremap}(D) + \xi}{|D| + \zeta} \triangleq \frac{\sum_{i=1}^n \featuremap(x_i) + \xi}{n + \zeta},
\end{equation}
where $\xi_j$ ($j = 1, ..., m$) and $\zeta$ are all Laplace random variables with scale parameter chosen according to \Cref{def:laplace-mechanism}.
For $\xi$, the scale depends on the sensitivity of the sum-of-features function, which can be expressed as $\Delta_1(\Sigma_{\featuremap}) = \max_x \|\featuremap(x)\|_1$, which can be computed as: $m' \sqrt{2}$ for RFF~\cite{chatalic2021CompressiveLearningPrivacy}, $R$ for RACE~\cite{coleman2020sub}, and $k$ for HIST~\cite{dwork2008differential}. For $\zeta$ the scale parameter depends on the sensitivity of the cardinality function which is always $\Delta_1(|\cdot|) = 1$.
The total privacy budget $\varepsilon$ is split across the numerator and the denominator, i.e. the noises $\xi$ and $\zeta$ are also respectively proportional to $\varepsilon^{-1}_{num}$ and $\varepsilon^{-1}_{den}$, with $\varepsilon = \varepsilon_{num} + \varepsilon_{den}$.
As stated above, such private sketches have already been considered in the literature and are not a contribution of this paper: we simply use sketches of this form in order to apply the $\mtom$ method introduced in the next section.

Although we focus on pure $\varepsilon$-DP in this manuscript for simplicity, private sketches can easily be extended to satisfy approximate DP (also known as $(\varepsilon,\delta)-$differential privacy) using the Gaussian mechanism~\cite{dwork2006calibrating}. This requires computing the $L_2$ sensitivity of the feature map, see for example~\cite{chatalic2021CompressiveLearningPrivacy,harder2021DPMERFDifferentiallyPrivate} for RFF.

\subsection{Related work}

The key advantages of sketching methods for data analysis with differential privacy is that they produce a private ``summary'' of the dataset, from which an arbitrary number of analyses can be performed. 
This idea of publishing a DP summary of the data has been explored in the literature, e.g., by Barak et al. with the release of full contingency tables~\cite{barak2007privacy}.
As contingency tables do not scale with the number of dimensions, further work has been proposed to publish so-called ``views'' of the data, from which $n-$way marginals can then be computed~\cite{qardaji2014priview}.
Another type of data summary that has gained popularity in recent years is \textit{synthetic data}, where the data curator publishes a dataset is ``similar'' to the original data, but with no mapping from real to synthetic records.
These usually involve training a statistical model on the data, which is then used to generate synthetic records, either explicitly~\cite{li2014differentially,zhang2017privbayes} or using generative networks~\cite{xie2018differentially}.

Kernel mean embeddings are known to carry a lot of information on the data distribution and are thus of particular interest for privacy applications.
Balog et al. suggested to use synthetic data points in order to represent (possibly infinite-dimensional) kernel mean embeddings in a private manner~\cite{balog2018}.
Finite-dimensional approximations based on random Fourier features have been made private using simple additive perturbation mechanisms with applications to clustering and Gaussian modeling~\cite{chatalic2021CompressiveLearningPrivacy} as well as synthetic data generation~\cite{harder2021DPMERFDifferentiallyPrivate}.
More recently, compact sketches based on Hermite polynomials have been proposed~\cite{park2021PolynomialMagicHermite} and have been shown empirically to provide a better privacy-utility tradeoff for private data generation than random Fourier features.

Relating specifically to the $\mtom$ method, the idea of considering a learned linear combination of random features (without privacy) has been popularized by Rahimi and Recht~\cite{rahimi2009weighted}, and then extensively studied under the name of “extreme learning machines” (ELMs)~\cite{huang2006extreme,huang2015TrendsExtremeLearning}.
The sketches considered in this paper can be interpreted as instances of this idea, with an additional averaging operation over the dataset. This is made possible by the fact that we only consider learning moments of the data.

\section{The moment-to-moment method}
\label{s:mtom}

Sketching methods can be used to efficiently perform specific learning tasks, and can often be made private in a straightforward manner by additive perturbation; however, extracting information from them is hard in general.
Here, we introduce the \textit{moment-to-moment} ($\mtom$) heuristic to learn a broad range of aggregate statistics from a single sketch. While previous sketched learning methods were relatively specific in the sense that both the feature map and the algorithm to learn from the sketch had to be tailored to a specific machine learning task, our heuristic can be used to approximate various kinds of statistics from the same sketch.
Although $\mtom$ can naturally be used on a non-private sketch, it is particularly attractive for private sketches, as it allows an analyst to perform arbitrarily many analyses from the sketches without incurring any additional privacy budget.

In the following, we assume that the data curator holds a sensitive dataset $D$ of size $n$, chooses a data-independent feature map $\featuremap$, and releases publicly the triplet $(\featuremap, \privatesketch{\dataset}, n+\zeta)$ where $\privatesketch{\dataset} = \frac{1}{n + \zeta} (\sum_{i=1}^n \featuremap(x_i) + \xi)$ is the private sketch computed as in \eqref{eq:private-sketch} and $\xi,\zeta$ are random and chosen as explained in \Cref{s:dp_sketching} in order to ensure $\varepsilon$-DP (i.e. they depend on the sensitivity of the feature map $\featuremap$).
Note that any result obtained by post-processing from this triplet will always remain $\varepsilon-$DP.

\subsection{Method description}

Suppose that an analyst wants to compute the empirical average of an arbitrary target function $f : \bb R^d \rightarrow \bb R$ over the dataset, i.e. the quantity $\overline{f} \triangleq \frac{1}{n}\sum_{i=1}^n f(x_i)$. The $\mtom$ method estimates $\overline{f}$ by a linear function of the sketch $\left<a,\privatesketch{D}\right>$, where the coefficients $a \in \mathbb{R}^m$ are computed by the method.
Because both the input (the sketch $\privatesketch{D}$) and the output (the target average $\overline{f}$) can be seen as ``generalized moments'' (averages of some features of the data) of the dataset, this amounts to transforming one type of generalized moment to another, hence the name of our method.

In order to apply the method, an analyst chooses \textit{a priori} a bounded domain $\domain \subset \bb R^d$ such that all possible records lie inside of $\domain$ (for example, $\domain$ might be a box constrained by physical upper and lower bounds on the data values).
The principle of $\mtom$ is to approximate the target function $f:\mathbb{R}^d \rightarrow \mathbb{R}$ over this domain $\domain$ by a linear model $\widetilde{f}$ of parameters $a \in \mathbb{R}^m$ in the output space of the sketch feature map $\featuremap : \bb R^d \rightarrow \bb R^m$, i.e.,
$$\widetilde{f}(x) \triangleq \langle a, \featuremap(x) \rangle \approx f(x), \quad \forall x \in \domain.$$
The key insight is that this linear model can then be used to estimate the dataset average $\overline{f}$ from the dataset sketch $\sketch{\dataset}$, since the sketching operator is linear:
\begin{equation}
\textstyle
    \widetilde{f}(\sketch{\dataset}) = \langle a, \sketch{\dataset} \rangle =\frac{1}{n} \sum_{i=1}^n \langle a, \featuremap(x_i) \rangle \approx \frac{1}{n}\sum_{i=1}^n f(x_i) = \overline{f}.
    \label{eqn:m2m}
\end{equation}

Intuitively, the target function $f(\cdot)$ is approximated by a linear combination $\langle a,\featuremap(\cdot) \rangle$ of a set of $m$ base functions (the components of the feature map, $\featuremap_i(\cdot)$). The quality of the approximation depends on the compatibility between the feature map $\featuremap$ and the function to approximate $f$.
Zhang et al.~\cite{zhang2012UniversalApproximationExtremeb} showed that such a linear combination can approximate continuous functions arbitrary well for a large enough number of features $m$, under conditions satisfied by many standard feature maps.
This suggests that $\mtom$ can be used to approximate any continuous function $f$ for a large array of sketches, although quantifying precisely how the approximation quality decreases with $m$ is out of the scope of this paper.
It should also be expected that approximating a discontinuous function $f$ with, e.g., RFF features will lead to high approximation error (e.g., some kind of Gibbs phenomenon).

\begin{figure}
    \centering
    \includegraphics[width=6cm, trim=1cm 1cm 0cm 1cm]{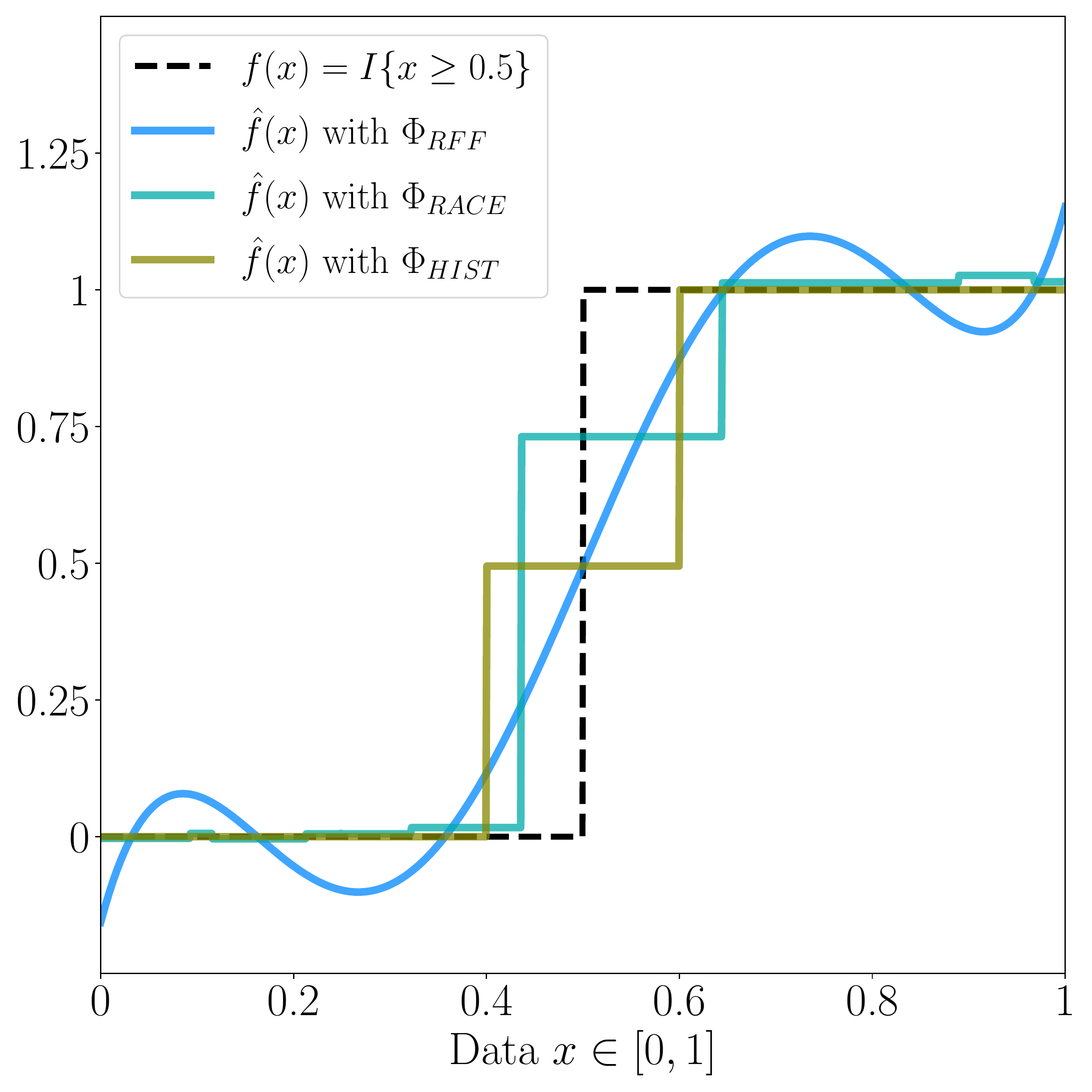}
    \caption{\textbf{How $\mtom$ works}: the $\mtom$ method approximates the target function $f$ as a linear combination of components of the feature map, $\sum_{i=1}^m a_i \featuremap_i(x) \approx f(x)$.
    }
    \label{fig:m2m-illustration}
\end{figure}

We illustrate $\mtom$ with a toy example in Fig.~\ref{fig:m2m-illustration}. We consider the step function $f(x) = I\{x\geq 0.5\}$ restricted to the domain $\domain = [0,1] \subset \mathbb{R}^{d=1}$.
For RFF, the approximation $\widetilde{f}(x)$ is a linear combination of $\cos(\omega_i^Tx)$ and $\sin(\omega_i^Tx)$, for some fixed $\omega_i$, which explains the bumps observed in the approximation (Gibbs phenomenon).
The RACE feature map, whose base functions are the one-hot encoding of locally-sensitive hash functions, approximates $f$ by a piecewise constant function.

\subsection{Optimizing the $\mtom$ model}
\label{sec:optimizing-m2m}

For the results produced by the $\mtom$ method to be useful, the parameters of the linear model $a$ need to be chosen such that $\widetilde{f}(\cdot)$ is a good approximation of the true function $f(\cdot)$ on the domain of interest.
For this, we formulate and optimize a loss function $J$ for the vector of weights $a$ that penalizes differences between $f$ and $\widetilde{f}$.
The full learning procedure is described in algorithm~\ref{algo:m2m} in appendix~\ref{app:m2m}.

Similarly to Rahimi et al.~\cite{rahimi2008random}, we use the squared difference as distance, $d(f(x), \widetilde{f}(x)) = (\tilde{f}(x) - f(x))^2$.
Assume that the records $x_i \in D$ are drawn from some (unknown) probability distribution $x_i \distiid p_X$.
Ideally, the $\mtom$ procedure would minimize the average error of the approximation \textit{over the true data distribution} $p_X$, 
$J_\text{ideal}(a) = \expectp{d\left(f(x), \widetilde{f}(x)\right)^2}{X\sim p_X}$.
However, the analyst only has access to the private sketch and does not know  $p_X$, let alone the data $D$. Instead, they choose an \textit{a priori} distribution $\psi$ that is either (1) close to $p_X$, or (2) likely to yield a good approximation where $p_X$ takes significant values when optimizing for the (approximate) loss $J_{\psi}(a) = \expectp{d\left(f(x),\widetilde{f}(x)\right)^2}{X \sim \psi}$.
In this work, we assume no prior knowledge except for the domain $\domain$ and thus use the uniform distribution on this domain $\psi = Unif(\domain)$, following the principle of maximum entropy.
Finally, since evaluating the expectation operator analytically can be challenging for arbitrary $\psi$, $f$ and $\Phi$, especially in high dimensions, we approximate it by sampling a large number $\nsynth$ of \textit{training synthetic} data points $\left(\tilde{x}_i\right)_{i=1}^{\nsynth}$ sampled i.i.d. from $\psi$.
The resulting loss, given a choice of $\psi$, is:
\[
\textstyle
J_\text{noreg}(a) = \frac{1}{N} \sum_{i=1}^N \left(f(X_i) - \left<a, \Phi(X_i)\right>\right)^2 ~~~ X_1, \dots, X_N \sim_\text{i.i.d.} \psi
\]
However, minimizing $J_\text{noreg}$ directly is not robust to noise, and in particular to the noise added to obtain differential privacy. Indeed, when applying the linear model $a$ from~\eqref{eqn:m2m} to the private data summary $\privatesketch{\dataset}$, we get (neglecting, for illustration, the noise $\zeta$ on the denominator):
\[
\textstyle
\langle \privatesketch{D}, a \rangle = \left<\frac{1}{n}\sum_{i=1}^n \Phi(x_i) + \xi, a\right> \approx \frac{1}{n} \sum_{i=1}^{n} {f}(x_i) + \frac{1}{n} \langle \xi, a \rangle.
\]
Hence, the noise on the numerator $\xi$ causes an error in the $\mtom$ estimate of variance $\sigma_{\xi}^2 \|a\|^2_2 / n^2$. To account for this noise, we add a term proportional to its variance to the loss function $J$:
\begin{equation}
\textstyle
    \label{eqn:mem_loss}
    J(a) \triangleq \expectp{\left(f(X) - \left<a, \featuremap(X)\right>\right)^2}{X\sim\psi} + \lambda \|a\|_2^2,
\end{equation}
where we set the regularization parameter $\lambda$ to the value $\sigma_{\xi}^2/ n^2$.
We prove that this loss $J$ is an upper bound for the mean square prediction error between $\bar{f}$ and the $\mtom$ estimate $\left<a, \privatesketch{D}\right>$ (see proof in Appendix~\ref{app:proof}).
\begin{theorem}
\label{thm:m2m-loss}
Let $\Phi:\mathbb{R}^d\rightarrow\mathbb{R}^m$ be a feature map, $\domain \subset \mathbb{R}^d$, and $\dataset$ be a random dataset of $n$ records $X_1,\dots,X_n \distiid \psi$. For all $a \in \mathbb{R}^m$, and all distributions $\psi$, if $\lambda = \sigma_{\xi}^2/n^2$ and $\zeta = 0$, we have that, if $\zeta=0$:
\[
\textstyle
J(a)
\geq
\expectp{\left(\frac{1}{n}\sum_{i=1}^n f(X_i) - \left<a, \privatesketch{\dataset}\right>\right)^2}{X_1,\dots,X_n,~\xi}
\]
\end{theorem}
Since the exact dataset size $n$ is not directly available to the analyst, we use $|D|+\zeta$ as an estimation of $n$. Further to this, we found empirically that using $\lambda = \frac{\sigma_{\xi}^2}{n^2}$ makes the model insufficiently robust to noise (especially when the sensitivity of the feature map is large). We thus use a larger regularization parameter in experiments by removing the square on the estimated number of samples.
\begin{equation}
\textstyle
    \lambda = \frac{\sigma_{\xi}^2}{\left(|D| + \zeta\right)} = \frac{2 \cdot \Delta_1(\featuremap)^2}{\varepsilon_{num}^2 \cdot \left(|D|+\zeta\right)}.
\end{equation}

\paragraph*{Solving for $J$}
Let $(\tilde{x}_i)_{i=1}^{\nsynth}$ denote the set of random training samples used inside the $\mtom$ procedure. Denote the synthetic feature matrix
$\mathbf{P} = \left(\featuremap(\tilde{x}_i)\right)_{i=1}^{\nsynth} \in \mathbb{R}^{\nsynth \times m}$, and the vector of corresponding outputs $\mathbf{F} = \left(f(\tilde{x}_i)\right)_{i=1}^{\nsynth} \in \mathbb{R}^n$. The empirical loss that $\mtom$ optimizes is $J(a) = \frac{1}{\nsynth}\|\mathbf{P}\cdot a - \mathbf{F}\|_2^2 + \lambda \|a\|_2^2$. This corresponds to a ridge regression problem with regularization parameter $\lambda$, and can be solved efficiently.

\subsection{Sources of error}

$\mtom$ is a heuristic method to approximate $\bar{f}$, and as such will always incur some error. We here outline the four main sources of error of $\mtom$.

\begin{enumerate}

    \item \textit{Sampling error}: The expectation operator in the cost function $J(a)$ is not computed exactly, but estimated by sampling $\nsynth$ points $\tilde{x}_i \sim \psi$. If $\nsynth$ is too small, this estimate can be inaccurate, and the model $a$ risks ``overfitting'' to the small training set.

    \item \textit{Approximation error}: $\mtom$ finds coefficients $a$ such that the linear combination $\widetilde{f}(\cdot) = \left<a,\featuremap(\cdot)\right> = \sum_{i=1}^m a_i \featuremap_i(\cdot)$ approximates the target function $f$. In general, even if $a$ is the exact minimizer of $J(a)$, there remains some inherent approximation error which depends on the compatibility between the feature map $\featuremap$ and target function $f$.

    \item \textit{Distributional shift}: In practice, the empirical distribution $p_X$ differs from the probability distribution~$\psi$ used for training. 
    Distributional shift is a hard problem to fix, as it requires tailoring $\psi$ to $p_X$ \textit{without accessing the data}, or only through the sketch. We discuss this in~\Cref{s:discussion}.

    \item \textit{Differential privacy noise}: Finally, the noises $\xi$ and $\zeta$ added in the computation of the sketch $\privatesketch{\dataset}$ further distort the representation. This error decreases when the privacy budget $\varepsilon$ increases.

\end{enumerate}

\subsection{Statistical estimation with $\mtom$}
\label{s:tasks_for_mtom}

Many learning tasks can be written as the estimation of some generalized moments of the data. Here we give some common examples.

\noindent\textbf{1. Moments}:
The $j^\text{th}$ component of the $k^\text{th}$ moment of the empirical data distribution is defined as
\[ \textstyle m^{(k)}_j = \frac{1}{n} \sum_{i=1}^n (x_i)_j^k \approx \mathbb{E}_{X\sim p_X} X_j^k, \]
which is the empirical average of the function $f^{(j, k)}:\mathbb{R}^d\rightarrow\mathbb{R}:x \mapsto x_j^k$.

\noindent\textbf{2. Counting queries:}
Given a set $S \subset \mathbb{R}^d$, a counting query over $\dataset$ consists of finding the number of data points from the dataset $\dataset$ that belong to~$S$:
\[ \textstyle \countq(\dataset,S) = \left|\left\{i\in \{1,\dots,n\}: D_i \in S\right\}\right| =\sum_{1\leq i\leq n} f_S(x_i). \]
where $f_S:\mathbb{R}^d \rightarrow \{0,1\}: x \mapsto I\{x\in S\}$ denotes the indicator function of $S$.
Histograms are a specific subset of counting queries, where the set $S$ is chosen to be a one-dimensional ``bin''.

\noindent\textbf{3. Covariance:}
The $(i,j)^\text{th}$ entry of the empirical covariance matrix is
\[
\textstyle
c_{ij} = \frac{1}{n}\sum_{l=1}^n ((x_l)_i - \mu_i) \cdot ((x_l)_j - \mu_j),
\]
which is the empirical average of the function $f^{(i,j)}:\mathbb{R}^d\rightarrow\mathbb{R}:x \mapsto (x_i-\mu_i)(x_j-\mu_j)$. The mean of the component $i$, $\mu_i$, can be estimated using $\mtom$ for the first-order moment, $m_i^{(1)}$.

\subsection{Classification and regression by approximation of the loss}
\label{subsec:im2m}

Many learning tasks can be formulated as learning a parametric model with parameter $\theta$ using a loss function $L$. For such tasks, one will typically solve the optimization problem
$\textstyle \theta^* \in \arg\min_{\theta}\textstyle \frac{1}{n} \sum_{i = 1}^n L(x_i,\theta)$, whose objective function takes the form of a generalized moment.
Specifically, for a classification or regression task, the analyst wants to fit some model $F_\theta : \bb R^{d-1} \rightarrow \bb R$ parameterized by $\theta \in \bb R^p$ to the data samples $(x_i)_{i=1}^n$, where each sample $x_i$ is a pair $x_i=(\overline{x}_i\in \bb R^{d-1}, y_i\in \bb R)$.
If the fitting quality is quantified by a loss function $l(.,.)$, one can define $L_{\theta}(x_i)\triangleq L(x_i,\theta) \triangleq  l\left(F_{\theta}(\overline{x}_i), y_i\right)$
and $\mtom$ can be used with the target $f=L_{\theta}$ for any \textit{fixed} value of $\theta$.
Finding the optimal parameter $\theta^*$ involves solving the following bi-level optimization problem:
\begin{equation}
    \theta^* \in \arg\min_\theta \left<a_\theta, \privatesketch{\dataset}\right> ~~ \text{such that} ~~ a_\theta \in \arg\min_a J_{\theta}(a)
    \label{eq:implicit-m2m-problem}
\end{equation}
where $J_{\theta}$ is the $\mtom$ objective associated to the target function $L_{\theta}$.
As mentioned in Section~\ref{sec:optimizing-m2m}, solving for $a$ is a ridge regression a problem, which has a closed-form solution (given synthetic samples $\tilde{x}$ used to compute $J$) of
$
a_\theta = \mathbf{S}\cdot\sum_{i=1}^{\nsynth} \Phi(\tilde{x}_i) L_\theta(\tilde{x}_i)
\quad \text{where} \quad
\mathbf{S} = \left(\frac{1}{\nsynth}\sum_{i=1}^{\nsynth}\Phi(\tilde{x}_i)^T\Phi(\tilde{x}_i) + \lambda I\right)^{-1}.
$
We then use this result in eq.~\ref{eq:implicit-m2m-problem} to formulate the dual optimization problem as an optimization problem in $\theta^*$:
\[
\theta^* \in \arg\min_\theta \sum_{i=1}^{\nsynth} \underbrace{\Phi(\tilde{x}_i)^T \cdot \mathbf{S} \cdot \privatesketch{\dataset}}_{\triangleq w(\tilde{x}_i)}~\cdot~ L_\theta(\tilde{x}_i)
\]
This method, which we call \textit{implicit-$\mtom$}, computes a weighting function $w:\Omega \rightarrow \mathbb{R}$ from the feature map $\featuremap$, private sketch $\privatesketch{\dataset}$, and regularization coefficient $\lambda$, independently of the loss.
This weighting function is used to weigh the contribution of each synthetic points to the total loss.
Any learning procedure, such as gradient descent, can then be applied to the re-weighted loss.

\section{Experiments}
\label{sec:results}

We empirically evaluate the $\mtom$ method on a range of data analysis tasks on artificial and real data. We perform our analyses on the LifeSci dataset, a real-world dataset of life sciences measurements ($n = 2.7 \cdot 10^4$ records and $d=10$ attributes), which we normalize to $\Omega = [0,1]^d$.
In order to analyze the different sources of errors independently, we perform the same analyses on a uniformly sampled artificial dataset of same shape ($n, d$), which we call \textsf{Random10}.
Since the training distribution $\psi$ is equal to the empirical distribution $p_X$, there is no \textit{distributional shift}, and the error observed in the results for \textsf{Random10} is thus the combination of \textit{approximation error}, \textit{sampling error} and \textit{the DP noise addition}.

We sketch each dataset the using RFF ($m = 200, \sigma = 1$), RACE ($R=W=80$), and HIST (marginals of each attribute, $n_{bins} = 100$ bins of same size in $[0,1]$), and add noise to ensure DP with privacy budget $\varepsilon \in [10^{-2}, 10^2]$, as described in~\Cref{s:dp_sketching}.
For all sketches, we split the privacy budget as $\varepsilon_{num} = 0.98\,\varepsilon$ and $\varepsilon_{den} = 0.02\,\varepsilon$.
We train $\mtom$ models with $\nsynth = 10^5$ samples, which empirically results in very low sampling error (training and testing $R^2$ scores almost identical). We repeat each experiment $50$ times and report, for each task, the average accuracy over all runs.

An alternative to sketches is synthetic data generation (SDG), where a statistical model is fit to the real data, and so-called synthetic data are then generated by sampling from this model.
We compare our results with datasets generated using three differentially private SDG methods: \textbf{DP-Copula}~\cite{li2014differentially}, \textbf{PrivBayes}~\cite{zhang2017privbayes}, and \textbf{DP-WGAN}~\cite{xie2018differentially}.
The latter method relies on a relaxed definition of differential privacy, $(\varepsilon,\delta)-$DP, and hence the guarantees provided are weaker. In our experiments, we use $\delta = 10^{-5}$.
For each SDG and $\varepsilon$, we generate $10$ synthetic datasets from LifeSci, and perform the tasks of interest on the synthetic data (by computing the empirical average of the functions $f$ on the synthetic data), reporting the average over all runs.

\subsection{Tasks involving columns in isolation}

\begin{figure}
    \centering
    \includegraphics[width=1.3\textwidth, trim=10cm 0 0cm 0]{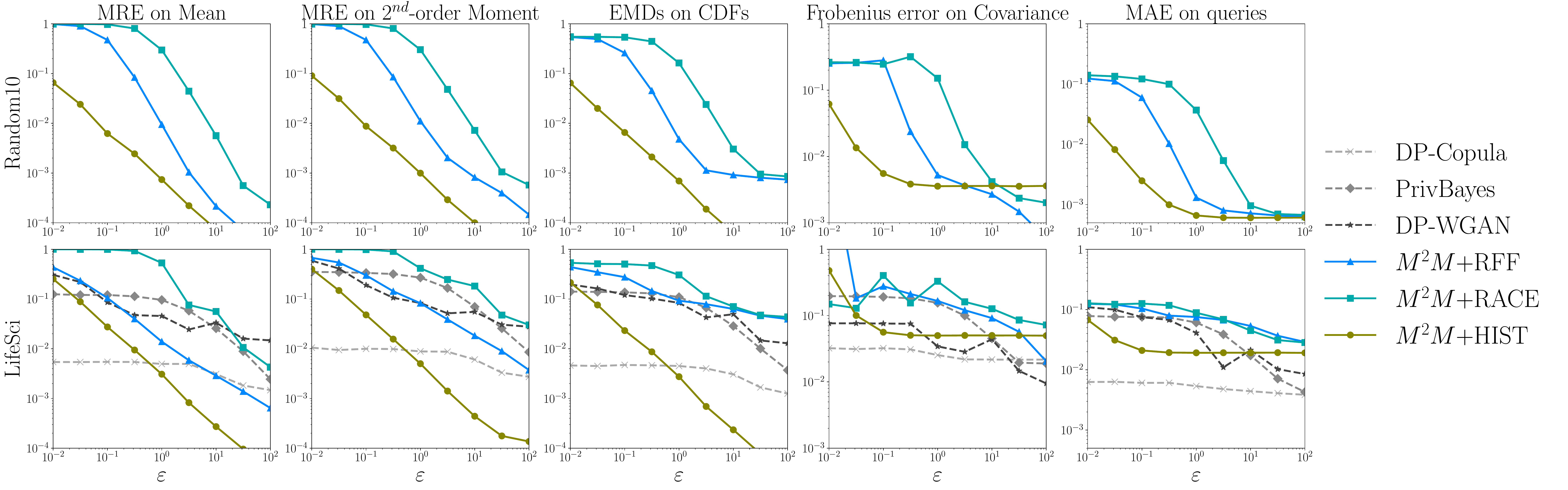}
    \caption{\textbf{Estimation of one-dimensional statistics over a random artificial dataset (top row) and LifeSci (bottom row)}. We estimate the mean, second-order moment, and CDF of each attribute using $\mtom$ on three sketches (RFF, RACE, and HIST), and synthetic datasets (generated using DP-Copula, PrivBayes, and DP-WGAN).
    We estimate the covariance matrix and the answer to a large number of random counting queries using $\mtom$ on three sketches (RFF, RACE, and HIST), and synthetic datasets (generated using DP-Copula, PrivBayes, and DP-WGAN).}
    \label{fig:m2m-results-combined}
\end{figure}

As a first illustrative example, we consider a range of simple tasks where the function learned with $\mtom$ only concerns one attribute in isolation.
For each sketch and each column in the datasets, we train a $\mtom$ model to predict (1) its mean $\frac{1}{n}\sum_{i=1}^n x_i$, (2) its order 2 moment $\frac{1}{n}\sum_{i=1}^n x_i^2$, and (3) its cumulative distribution function (CDF) in 10 equi-distant points $\left(\frac{1}{n}\sum_{i=1}^n I\{x_i \leq S_j\}\right)_{j=1}^{10}$.
We then measure the error obtained between the predicted value and the empirical value using mean relative error (MRE) $MRE(\hat{\mu}, \mu) = \frac{\left|\hat{\mu} - \mu\right|}{\mu}$ for (1) and (2), and the Earth-Mover Distance (EMD) for (3)
For each task, sketch, and dataset, we report the average error across all attributes in Fig.~\ref{fig:m2m-results-combined}.

We show that, in the absence of distributional shift, $\mtom$ can be used to estimate single-variable tasks with good accuracy. As expected, the HIST sketch performs well on all tasks and for both datasets, since it is specifically designed to approximate one-dimensional distributions. However, distributional shift (in LifeSci) worsens results significantly for all feature maps.
This is particularly true for CDF, where the RFF and RACE feature maps result in high error, probably due to high approximation error.
Comparing with synthetic data, we find that the RFF sketch compares favorably with both PrivBayes and DP-WGAN (especially when $\varepsilon \geq 1$), while the RACE sketch leads to less useful results.
DP-copula datasets outperform both sketches, which is to be expected since the method explicitly estimates marginals.

We further analyze the different sources of error in table~\ref{tab:different_errors}. We report the mean relative error on the first moment $\mathbb{E}[X]$ obtained with either the exact sketch ($\varepsilon=+\infty$) or the private sketch with parameter $\varepsilon=1$, for the RFF and HIST feature maps, on both datasets.
For the HIST feature map and $\varepsilon=+\infty$, we find the $\mtom$ coefficients using a small regularization $\lambda = 10^{-9}$ (for numeric stability).
The error on the artificial dataset for $\varepsilon=+\infty$ is the \textit{approximation error} of $f$, the irreducible error obtained when approximating $f$ by a linear mixture of components of the feature map $\featuremap_i$.
We observe that this error is low for the RFF feature map, which has strong approximation properties~\cite{rahimi2008random,zhang2012UniversalApproximationExtremeb}, and higher for the HIST sketch, which roughly approximates a function as a product of 1D piecewise constant functions.
The second row ($\varepsilon=1$) is the result of adding \textit{DP error} to the approximation error. DP error has a negligible impact on the performances of the histogram sketch, as it is dominated by the approximation error. The opposite applies to RFF, where the DP error is 5 orders of magnitude larger.
Results from the LifeSci dataset (rows 3 and 4) illustrate the impact of distributional shift, when the distribution used to generate $\mtom$'s training set differs from the empirical distribution.
For $\varepsilon=1$, we observe that all resulting errors are one order of magnitude larger, as a result of distributional shift.
Furthermore, as expected, when there is no DP error ($\varepsilon=+\infty$), the approximation error for LifeSci is higher than for the Random10, for both sketches.
Hence, distributional shift can have disparate effects on the resulting accuracy of the method, by amplifying either or both of the approximation and DP error.

\begin{table}
    \centering
    \begin{tabular}{l|l|c|c}
        Dataset & Budget & MRE for $\featuremap^\RFF$ & MRE for $\featuremap^\text{HIST}$ \\ \hline
        Random10~ &$\varepsilon=+\infty$ & $6.25 \cdot 10^{-8}$ & $1.87\cdot 10^{-5}$\\
         & $\varepsilon=1$ & $9.55 \cdot 10^{-3}$ & $9.10\cdot 10^{-4}$\\ \hline
        LifeSci &$\varepsilon=+\infty$ & $1.67\cdot 10^{-6}$ & $5.91\cdot 10^{-5}$ \\
         & $\varepsilon=1$ & $4.20 \cdot 10^{-2}$ & $3.8\cdot 10^{-3}$ \\
    \end{tabular}
    \caption{\textbf{Comparison of asymptotic, DP, and distributional shift errors}: We measure the RMSE on the first moment $\mathbb{E}[X]$ estimated with the $\mtom$ method and the Random Fourier Features $\featuremap^\RFF$ and HIST $\featuremap^\text{HIST}$ feature maps, on the artificial and LifeSci datasets. We report the asymptotic error (no noise) and the error for $\varepsilon=1$. All results are averaged over~100 trials.}
    \label{tab:different_errors}
\end{table}

\subsection{Multi-column tasks}

We evaluate $\mtom$ on tasks that involve attributes taken together. First, we compute the covariance matrix of the dataset, $\frac{1}{n}\left((x_i - \hat{\mu}_i)\cdot (x_j - \hat{\mu_j})\right)_{i=1, j=1}^{n, n}$, using $\hat{\mu}_i$ estimated as above.
We measure the Frobenius distance between the estimated and empirical covariance matrices.
Next, we perform a large number of simple counting queries $\countq(D, S)$, where the query $S$ is defined as the conjunction of three predicates of the form $X_i \leq u$ or $X \geq l$, for three different attributes $X_i, X_{i'}, X_{i"}$. We report the Mean Absolute Error (MAE) between the real query answers and the answers predicted by $\mtom$.

Fig.~\ref{fig:m2m-results-combined} reports the error decrease for both tasks and on each dataset as $\varepsilon$ increases.
Similarly to the one-dimensional tasks (Fig.~\ref{fig:m2m-results-combined}), we observe that $\mtom$ estimations perform well on the Random10 dataset, and worse on LifeSci.
Except for PrivBayes, all synthetic datasets (and in particular, DP-Copula) outperform $\mtom$.
The queries use case is particularly challenging to approximate with $\mtom$, as the target function $f$ is not continuous.
Finally, as expected, results for the HIST sketch quickly plateau for all tasks and datasets.

\subsection{Logistic Regression}

We use the implicit-$\mtom$ method described in section~\ref{subsec:im2m} to perform logistic regression from the private sketch of a dataset.
We use real-world building occupancy data~\cite{candanedo2016accurate} ($d=6$, $n=20,560$) with $5$ continuous attributes (building characteristics) and a binary attribute (whether a building is occupied).
This dataset is such that the last attribute is strongly predicted by the continuous attributes, with an AUC (area under curve) of $> 0.99$.
We normalize the continuous attributes to $[0,1]$ and define $\Omega = [0,1]^5 \times \{0,1\}$ and $\psi = Unif(\Omega)$.
We randomly separate the data between training (90\%) and testing (10\%), then sketch the training dataset using RFF ($\sigma = 1, m = 200$), RACE ($R = 80, H = 80, \sigma = 0.1$) and HIST ($n_{bins} = 20$) for a range of $\varepsilon$.
Using implicit-$\mtom$, we perform logistic regression on each sketch and evaluate the result on the testing dataset.
We compare our results with Chaudhuri et al.~'s DP-ERM~\cite{chaudhuri2009privacy}, a dedicated method to train a logistic regression with DP using objective perturbation.

We also generate synthetic datasets using the same SDG techniques as above.
We train a logistic regression using \textsf{sklearn} on each dataset $10$ times, and measure its AUC on the test dataset.
It can happen that the synthetic dataset only has one class for the last attribute; in this case we report the AUC to be $0.5$.

\begin{figure}
    \centering
    \includegraphics[width=0.5\textwidth]{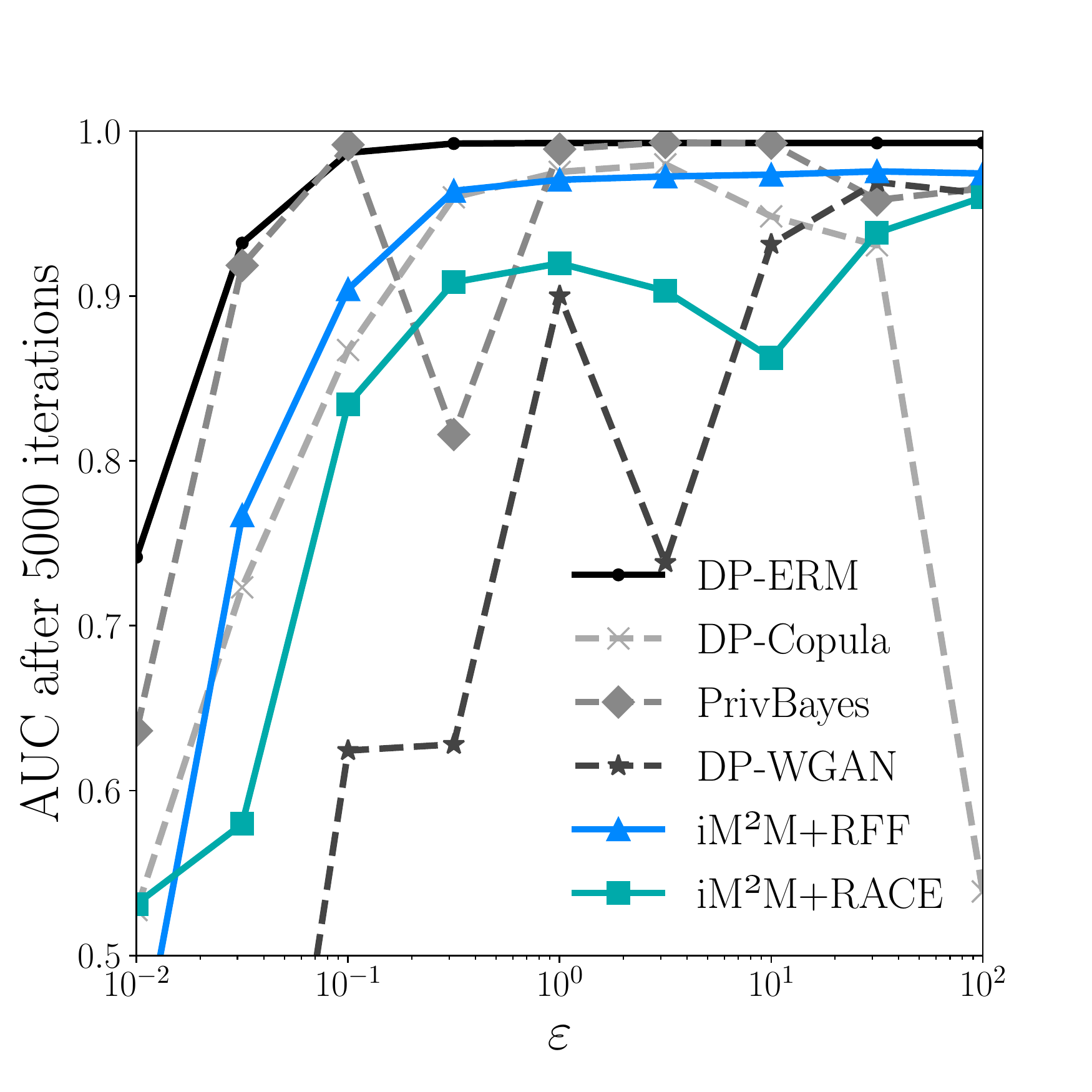}
    \caption{\textbf{AUC of a logistic regression trained from sketches on the occupancy dataset.}
    We use implicit-$\mtom$ to fit a logistic regression to the \textsf{occupancy} dataset from RFF and RACE sketches. We compare our results with the dedicated method DP-ERM and three synthetic data generation methods.}
    \label{fig:im2m-occupancy-logistic}
\end{figure}

In Fig.~\ref{fig:im2m-occupancy-logistic}, we show that implicit-$\mtom$ compares remarkably well with DP-ERM for the RFF feature map.
While it leads to higher erreor, the RACE feature map consistently produces an AUC of at least $0.9$ for $\epsilon \geq 0.3$.
Unsurprisingly, the method performs poorly on the HIST feature map ($AUC < 0.1$, not featured on the plot), which cannot, by definition, be used to estimate correlations between attributes.
Importantly, models trained with implicit-$\mtom$ compare favorably with models trained on synthetic datasets using the same budget $\varepsilon$.
As expected, the task-specific DP-ERM outperforms all other methods, but this comes at the cost of the entire budget $\varepsilon$.
Our results suggest that implicit-$\mtom$ is a promising solution to perform sophisticated learning tasks on sketches.

\section{Future work and conclusion}
\label{s:discussion}

Distributional shift occurs when the distribution used to generate $\mtom$'s training set, $\psi$, differs from the data distribution $p_X$. This is a significant source of error in the method. We here propose a few options to reduce this error.

\begin{itemize}

    \item Improving the approximation $\psi \approx p_X$ using the sketch. KDE sketches are built to approximate a kernel, encoding a kernel density estimate for the data distribution: $p_X(x) \approx \frac{1}{n} \sum_{i=1}^n \kappa(x,x_i) \approx \left<\featuremap(x),\privatesketch{\dataset}\right>$. One could thus use $\psi: x\mapsto\left<\featuremap(x),\privatesketch{\dataset}\right>$. However, the approximate distribution $\left<\featuremap(x),\privatesketch{\dataset}\right>$ can be negative and is not robust to noise addition for privacy.
    
    \item Learning a generative model on the sketch~\cite{harder2021DPMERFDifferentiallyPrivate} that, if accurately trained, generates synthetic data similar to the sketched dataset. These synthetic records can then be used to train the $\mtom$ model, as their distribution $p_\text{synth}$ is likely to be close to $p_X$ (or at least closer than $\psi$ uniform). Although the synthetic records could be used directly for the learning tasks, re-accessing the data sketch through the $\mtom$ mechanism could yield greater utility.

    \item Solving the loss minimization problem on the real data using a differentially private procedure. For instance, techniques such as DP-Empirical Risk Minimisation (DP-ERM)~\cite{chaudhuri2011differentially} could be applied -- although this can be challenging, since $J$ is non-convex. While this method is most likely the best solution to distributional shift, it requires additional privacy budget to learn the parameters of $\mtom$, which contradicts the idea of data summaries.

\end{itemize}

\bibliographystyle{splncs04}
\bibliography{bibliography}

\appendix

\section{Proof of Theorem~\ref{thm:m2m-loss}}
\label{app:proof}

Let $J_\Sigma$, the left-hand side of the inequality, the mean squared error between the empirical mean $\overline{f}$ and the estimation from the sketch $\widetilde{f}$. Denoting $X=(X_1,\dots,X_n)$, we have
\[
{
\footnotesize
\begin{array}{ll}
    J_\Sigma
    &= \expectp{\left(\frac{1}{n}\sum_{i=1}^n f(X_i) - \left<a, \frac{1}{n}\left(\sum_{i=1}^n \featuremap(X_i) + \xi\right)\right>\right)^2}{X,\xi}\\
    &= \expectp{\left(\frac{1}{n}\sum_{i=1}^n \left(f(X_i)-\left<a,\featuremap(X_i)\right>\right) - \frac{1}{n}\left<a, \xi\right>\right)^2}{X,\xi}\\
    &\stackrel{(i)}{=} \expectp{\left(\frac{1}{n}\sum_{i=1}^n \left(f(X_i)-\left<a,\featuremap(X_i)\right>\right)\right)^2}{X} + \frac{1}{n^2}\expectp{\left<a,\xi\right>^2}{\xi}\\
    &\stackrel{(ii)}{=} \frac{n (n-1)}{n^2} \cdot \left(\expectp{f(X)}{X}-\left<a,\expectp{\featuremap(X)}{X}\right>\right)^2\vspace{2pt}\\
    &\quad\quad~+~\frac{n}{n^2}\cdot\expectp{\left(f(X) - \left<a,\featuremap(X)\right>\right)^2}{X} + ||a||_2^2 \frac{\mathbb{V}[\xi]}{n^2}
\end{array}
}
\]
where we used in $(i)$ the independence from $\xi$ and $X$ and the fact that $\expect{\xi} = 0$, and in $(ii)$ the fact that samples $(X_i)_{1\leq i\leq n}$ are independent (and $\mathbb{V}[\cdot]$ denotes the variance of a random variable). 
Finally, we use Jensen's inequality (since $x\mapsto x^2$ is convex) to show that $\left(\expectp{f(X)}{X}-\left<a,\expectp{\featuremap(X)}{X}\right>\right)^2\vspace{2pt} \leq \expectp{\left(f(X) - \left<a,\featuremap(X)\right>\right)^2}{X}$, which concludes the proof.

\section{$\mtom$ learning procedure}
\label{app:m2m}

\SetKwInput{KwData}{Input}
\SetKwInput{KwResult}{Output}
\begin{algorithm}[h!]
 \KwData{Target function $f$, private data sketch $(\privatesketch{\dataset}, n+\zeta)$ with associated feature map $\featuremap$ and noise level $\sigma_\xi^2$, a priori distribution $\psi$, number of synthetic samples $\nsynth$, (optional) additional regularization $R=1$.}
 \KwResult{$\hat{f}$, an estimate for $\frac{1}{n}\sum_{i=1}^n f(x_i)$.}
Get $\nsynth$ synthetic training samples $\tilde{x}_i \distiid \psi$\;
Set regularization parameter $\lambda = \sigma_{\xi}^2/\left(|D| + \zeta\right)\cdot R$\;
Get coefficients $a = \arg\min_{\alpha}~\loss(\alpha)$, using the samples $\tilde{x}_i$ as estimation for $\psi$\;
\KwRet{$\hat{f} = \left<a,\privatesketch{D}\right> \approx \frac{1}{n}\sum_{i=1}^n f(x_i)$.}
 \caption{\textbf{$\mtom$ learning procedure}: Given a dataset sketch $\privatesketch{\dataset}$ and a target function $f:\mathbb{R}^d \rightarrow \mathbb{R}$, the procedure estimates the empirical mean of $f$ over $\dataset$.}
 \label{algo:m2m}
\end{algorithm}

\end{document}